\documentclass[aps,onecolumn,showpacs,footinbib]{revtex4}%
\usepackage[latin9]{inputenc}
\usepackage{amssymb}
\usepackage{amsmath}
\usepackage{amscd}
\usepackage{graphicx}
\usepackage{subfigure}
\usepackage{float}
\begin{document}
\draft
\title{Open system geometric phase based on system-reservoir joint state evolution}
\author{Shi-Biao Zheng\thanks{%
E-mail: sbzheng11@163.com}} \email{sbzheng11@163.com}
\address{Department of Physics,
Fuzhou University, Fuzhou 350116, China}
\date{\today }

\begin{abstract}
The geometric phase is of fundamental interest and plays an important role
in quantum information processing. However, the definition and calculation
of this phase for open systems remains a problem due to the lack of
agreement on generalizations of the parallel transport condition to mixed
state nonunity evolutions. Here we tackle this problem by associating the
open system geometric phase with the parallel transport of the joint
system-reservoir state. Our approach not only provides a way around the
nonunitary evolution obstacle, but also sheds light on the relation between
the geometric phase and the system-reservoir entanglement, which has not
been investigated. Based on this approach, we calculate the geometric phase
of different quantum systems subject to energy decay, showing that it is
robust against decoherence, which is in distinct contrast with previous
results.
\end{abstract}


\vskip 0.5cm\maketitle

\narrowtext

When the parameters of the Hamiltonian of a quantum system initially in an
eigenstate of the Hamiltonian is changed in a cyclic and adiabatic fashion,
it will return to the initial state but acquires a geometric phase in
addition to the dynamical one [1,2]. This effect, known as the Berry phase,
has attracted interest from a variety of fields [3] and been generalized in
various ways. Aharonov and Anandan removed the adiabatic condition and
showed that the geometric phase depends upon the geometry of the path
traversed by the system in the projected Hilbert space [4]. Based on the
work of Pancharatnam on the interference of polarized light [5], the
geometric phase associated with a noncyclic evolution was discovered [6].
The generalization to mixed quantum states was first introduced by Uhlmann
[7] and then developed by Sjoqvist et al., who defined the mixed state
geometric phase due to unitary evolution based on interferometry [8]. In
addition to fundamental interest, geometric phases have practical
applications, among which the implementation of geometric quantum
computation is a typical example. Quantum gates based on geometric phase may
have intrinsic resistance to external noise, offering potential advantages
as compared with dynamical ones [9-16].

One of the main obstacles for coherent control of quantum systems and
implementation of quantum computation is decoherence due to the presence of
noises, including fluctuations in the control parameters and unavoidable
interaction with the environment. This leads to many theoretical and
experimental studies of geometric phases under classical [17-24] and quantum
noises [25-35]. For the latter case, one should deal with the geometric
phase in an open system, whose definition is still an open problem though
approaches have been developed for this purpose [25-28]. For different
approaches the parallel transport conditions used to define mixed state
geometric phases are not consistent [26,36]. In Ref. [27], Carollo et al.
defined and calculated the geometric phase of a quantum system subject to
docoherence through the quantum jump method. It was shown that the geometric
phase is completely insensitive to dephasing but it is affected by the
spontaneous decay. In this framework, Carollo et al. investigated the
behavior of the geometric phase of a spin-1/2 particle interacting with a
decohering quantum field, showing that to the first correction this phase is
insensitive to decoherence for adiabatic evolution but is affected by
decoherence for nonadiabatic evolution [28]. This approach avoids the
problem of finding the parallel transport condition for mixed states, but it
requires continuous observation of the reservoir state, which would
interrupt the joint system-reservoir state evolution, destroying the
entanglement between the system and environment. Furthermore, it is
experimentally demanding to accurately detect whether a quantum jump has
occurred or not. Another problem is that for any evolution trajectory
involving one or more energy decay jumps the geometric phase cannot be well
defined since in this case the final and initial states become orthogonal.

The aim of the present paper is to overcome these problems and to present a
new method to define and calculate the geometric phase of quantum systems
coupled to reservoirs. Our method, based on the evolution of the
system-reservoir joint state, requires neither observation of quantum jumps
nor tracing over the degrees of freedom of the reservoir. Since the
reservoir state is included, the total system undergoes a unitary evolution
and the parallel transport condition is well defined. Within this framework,
the geometric phase is unique and depends upon the geometry of the path
traversed by the joint system-environment state in the total Hilbert space.
We show that this phase can be calculated without knowledge of the state
components with one or more excitations having leaked to the reservoir.
Based on this approach, we investigate the behavior of the geometric phase
of different open systems, showing that this phase is robust against
spontaneous decay. We also present a geometric explanation to this effect in
view of the correlated system-environment state evolution. This interesting
result is contrary to the previous study [26-28], which showed that the
geometric phase for nonadiabatic evolution has no resistance to system
decay. The reason for this difference is that in the previous works the
system-environment entanglement is lost either by measuring the system decay
or by tracing over the environmental degrees of freedom. We further show
that for the Jaynes-Cummings (JC) model, the fundamental model describing
the matter-light interaction at the quantum mechanical level, the open
system geometric phase calculated through our approach is the only one that
can be directly measured in experiment.

We consider a quantum system, whose Hamiltonian is given by $H_s$. Suppose
that it is coupled to an $N$-mode reservoir. The total Hamiltonian is given
by
\begin{equation}
H_{sr}=H_s+H_r+H_i,
\end{equation}
where $H_r=\hbar \sum_{k=1}^N\omega _kb_k^{\dagger }b_k$ is the reservoir
Hamiltonian, $H_i=\hbar \sum_{k=1}^Ng_k\left( b_k^{\dagger }o+b_ko^{\dagger
}\right) $ describes the system-reservoir interaction, and $b_k^{\dagger }$
and $b_k$ are the creation and annihilation operators for the $k$th
reservoir mode with the frequency $\omega _k$, which is coupled to the
system with the strength $g_k$. Here $o$ ($o^{\dagger }$) is the linear
combination of the relevant lowering (raising) operators for the energy
levels of the quantum system. Assume that both the system and reservoir are
initially in pure states so that their evolution follows a definite
trajectory $\left| \psi ^{sr}(t)\right\rangle $ in the total Hilbert space
of joint system-reservoir state under the total Hamiltonian $H_{sr}$. For
nonvanishing system-reservoir entanglement, the total geometric phase
acquired by the joint state evolution is inseparable. To give the parallel
transport condition associated with this phase, we define the new state
\begin{equation}
\left| \Phi ^{sr}(t)\right\rangle =e^{-i\phi _d}\left| \psi
^{sr}(t)\right\rangle ,
\end{equation}
where $\phi _d=-\int_0^t\left\langle \psi ^{sr}(t^{^{\prime }})\right|
H_{sr}\left| \psi ^{sr}(t^{^{\prime }})\right\rangle dt^{^{\prime }}$ is the
dynamical phase for the total system. With the dynamical phase being
removed, $\left| \Phi ^{sr}(t)\right\rangle $ satisfies the parallel
transport condition
\begin{equation}
\left\langle \Phi ^{sr}(t)\right| \frac d{dt}\left| \Phi
^{sr}(t)\right\rangle =0.
\end{equation}
After a time $T$ the geometric phase of the total system associated with
this parallel transport is
\begin{eqnarray}
\beta &=&\arg \left\langle \Phi ^{sr}(0)\right| \left. \Phi
^{sr}(T)\right\rangle  \nonumber \\
\ &=&\phi -\phi _d,
\end{eqnarray}
where $\phi =\arg \left\langle \psi ^{sr}(0)\right| \left. \psi
^{sr}(T)\right\rangle $ is the total phase difference between the initial
and final states $\left| \psi ^{sr}(0)\right\rangle $ and $\left| \psi
^{sr}(T)\right\rangle $. The geometric phase defined in this way, acting as
a memory of the joint motion of the system and environment, depends purely
upon the geometry of the trajectory followed by the joint system-reservoir
state in the total Hilbert space.

Suppose the system is initially in the state $\left| \psi
_0^s(0)\right\rangle $ and the reservoir initially in the vacuum state $%
\left| \stackrel{-}{0}\right\rangle =\prod_{k=1}^N\left| 0_k\right\rangle $.
The evolution of the total system is given by
\begin{equation}
\left| \psi ^{sr}(t)\right\rangle =\left| \psi _0^s(t)\right\rangle \left|
\stackrel{-}{0}\right\rangle +\left| \psi _1^{sr}(t)\right\rangle ,
\end{equation}
where $\left| \psi _1^{sr}(t)\right\rangle $ is the combined
system-reservoir state associated with one or more excitations being
transferred to the reservoir so that it is orthogonal to $\left| \psi
_0^s(t)\right\rangle \left| \stackrel{-}{0}\right\rangle $, which
corresponds to no excitation transfer. The evolution of $\left| \psi
_0^s(t)\right\rangle $ is governed by the conditional Hamiltonian $%
H_{s,c}=H_s-i\hbar \frac \gamma 2o^{\dagger }o$, where $\gamma $ is the
system energy decaying rate. Here $\left| \psi _0^s(t)\right\rangle $ and $%
\left| \psi _1^{sr}(t)\right\rangle $ are the unnormalized states, with $%
\left\langle \psi _0^s(t)\right| \left. \psi _0^s(t)\right\rangle $ being
the probability with no excitation decaying to the reservoir and $%
\left\langle \psi _1^{sr}(t)\right| \left. \psi _1^{sr}(t)\right\rangle $
the probability with one or more excitations in the reservoir modes.

When the system Hamiltonian is time independent, the total Hamiltonian is a
conserved quantity and
\begin{equation}
\phi _d=-\left\langle \psi ^{sr}(0)\right| H_{sr}\left| \psi
^{sr}(0)\right\rangle T.
\end{equation}
For the initially vacuum reservoir, we have $\left\langle \psi
^{sr}(0)\right| H_r\left| \psi ^{sr}(0)\right\rangle =\left\langle \psi
^{sr}(0)\right| H_i\left| \psi ^{sr}(0)\right\rangle =0$ and hence $\phi
_d=-\left\langle \psi ^{sr}(0)\right| H_s\left| \psi ^{sr}(0)\right\rangle T$%
. Since $\left\langle \psi ^{sr}(0)\right| \left. \psi
_1^{sr}(t)\right\rangle =0$, the total phase difference between the initial
and final states reducs to $\phi =\arg \left\langle \psi _0^s(0)\right|
\left. \psi _0^s(T)\right\rangle $. This implies that the geometric phase is
indepenent of expression of $\left| \psi _1^{sr}(t)\right\rangle $, which
makes the calculation of the geometric phase very simple. It should be noted
that this geometric phase can be directly measured in the system
interference experiment although it is obtained by the system-reservoir
joint state [37]. The total phase obtained by the present approach is the
same as that calculated through the quantum jump method for the no-jump
trajectory [27,28]. The dynamical phase calculated through the quantum jump
method is given by
\begin{equation}
\phi _d^{^{\prime }}=-\int_0^T\frac{\left\langle \psi _0^s(t)\right|
H_s\left| \psi _0^s(t)\right\rangle }{\left\langle \psi _0^s(t)\right|
\left. \psi _0^s(t)\right\rangle }dt.
\end{equation}
Since the evolution $\left| \psi _0^s(t)\right\rangle $ is governed by the
non-Hermitian Hamiltonian $H_{s,c}$, the system Hamiltonian $H_s$ is not a
conserved quantity for no-jump trajectory $\left| \psi _0^s(t)\right\rangle $
and hence $\phi _d^{^{\prime }}\neq \phi _d$. This implies that the
geometric phases obtained through these two approaches are different. This
difference has a simple explanation. In the present method the geometric
phase is associated with the real joint state evolution trajectory in the
total Hilbert space, while in the quantum jump method it depends upon the
projection of evolution trajectory to the subspace with no excitation in the
reservoir modes. In other words, observation of the environment destroys the
system-environment entanglement, changing the state trajectory of the total
system and, therefore, the related geometric phase. It should be noted that
for the no-jump trajectory the reservoir does not evolve and acquires no
geometric phase, and therefore the geometric phase acquired by the system is
also that acquired by the total system including the reservoir.

Let us analyze the geometric phases of typical physical systems based on
this approach. We first consider the evolution of a two-level atom in a
reservoir. This atom dispersively interacts with a classical electromagnetic
field, which does not induce the atomic transition, but shifts the atomic
energy levels. We here works in the framework rotating at the transition
frequency $\omega _a$ of the atom without energy shift. Then the system
Hamiltonian is given by $H_s=\frac 12\hbar B(\left| e\right\rangle
\left\langle e\right| -\left| g\right\rangle \left\langle g\right| )$, where
$\left| e\right\rangle $ and $\left| g\right\rangle $ are the upper and
lower levels of the atom and $B$ characterizes the relative energy level
shift caused by the off-resonant classical field. The system-reservoir
Hamiltonian has the form of Eq. (1), with the frequency $\omega _k$ of the $%
k $th reservoir mode being replaced by the detuning $\delta _k=\omega
_k-\omega _a$. Suppose the system is initially in the state $\left| \psi
_0^s(0)\right\rangle =\cos (\theta /2)\left| e\right\rangle +\sin (\theta
/2)\left| g\right\rangle )$. When the reservoir modes are closely spaced in
frequency, the system-reservoir state evolution is given by Eq. (5), where
[38]
\begin{eqnarray}
\left| \psi _0^s(t)\right\rangle &=&e^{-\gamma t/2-i\varphi /2}\cos (\theta
/2)\left| e\right\rangle +e^{i\varphi /2}\sin (\theta /2)\left|
g\right\rangle ,  \nonumber \\
\left| \psi _1^{sr}(t)\right\rangle &=&e^{-i\varphi /2}\sqrt{1-e^{-\gamma t}}%
\cos (\theta /2)\left| g\right\rangle \left| \stackrel{-}{1}\right\rangle ,
\end{eqnarray}
$\varphi =Bt$, $\left| \stackrel{-}{1}\right\rangle =\sum_{k=1}^N\lambda
_k\left| 0_1...0_{k-1}1_k0_{k+1}...0_N\right\rangle $ is the normalized
collective state for the reservoir modes with one excitation, and $\gamma $
is the decaying rate of the atom due to interaction with the reservoir.
Since the total Hamiltonian is conserved during the evolution, we have $%
\left\langle \stackrel{-}{1}\right| \left. \stackrel{.}{\stackrel{-}{1}}%
\right\rangle =0$. The total phase difference between the initial and final
states is $\phi =\arg \left[ e^{-\gamma T/2-iBT/2}\cos ^2(\theta
/2)+e^{iBT/2}\sin ^2(\theta /2)\right] $, while dynamical phase accumulated
by the overall system is $\phi _d=-\frac 12BT\cos \theta $. Remarkably, for $%
BT=2\pi $ (cyclic evolution in the absence of decoherence) the geometric
phase of the total system is $\beta =-\pi (1-\cos \theta )$, which is
completely unaffected by decoherence. For $\theta =\pi /2$, the dynamical
phase accumulated by the overall system is zero and the phase $\phi $ is of
purely geometric origin.

To interpret this result, we rewrite the state vector of the whole system as
\begin{equation}
\left| \psi ^{sr}(t)\right\rangle =e^{-i\varphi /2}\left[ \cos (\theta
/2)\left| \stackrel{-}{e}\right\rangle +e^{i\varphi }\sin (\theta /2)\left|
\stackrel{-}{g}\right\rangle \right] ,
\end{equation}
where $\left| \stackrel{-}{e}\right\rangle =e^{-\gamma t/2}\left|
e\right\rangle \left| \stackrel{-}{0}\right\rangle +\sqrt{1-e^{-\gamma t}}%
\left| g\right\rangle \left| \stackrel{-}{1}\right\rangle $ and $\left|
\stackrel{-}{g}\right\rangle =\left| g\right\rangle \left| \stackrel{-}{0}%
\right\rangle $. The variation of the state $\left| \stackrel{-}{e}%
\right\rangle $ with time does not produce any the phase to the state $%
\left| e\right\rangle \left| \stackrel{-}{0}\right\rangle $ and has no
effect on the acquired geometric phase $\beta $, which is determined by the
evolution of the relative probability amplitude of the two joint states $%
\left| \stackrel{-}{e}\right\rangle $ and $\left| \stackrel{-}{g}%
\right\rangle $. This allow us to illustrate the geometric aspect of the
joint state evolution with the Bloch sphere, whose north pole represents the
single-excitation state (the probability of the excitation being held by the
atom is irrelevant). With this representation, the coherent evolution
corresponds to a rotation of the Bloch vector around the z axis, and after a
complete cycle the system acquires a geometric phase equal to the solid
angel subtended by the path followed by vector always pointing to $(\theta
,\varphi )$ on the Bloch sphere. Here the system-reservoir entanglement
protects the coherence between the single- and zero-excitation states and
hence the geometric phase against the energy leakage.

It is interesting to further illustrate the difference between the present
result and that obtained through the quantum jump method with this typical
system. According to the quantum jump approach, the geometric phase
associated with the no-jump trajectory tends to $\beta -\gamma (\pi \sin
\theta )^2/(2B)$ up to the first order in $\gamma /B$. As pointed out in
[27], when the environment is monitored and no decaying is observed, the
probability of the lower state increases with time so that the state will
spiral towards the south pole in the Bloch sphere, changing the area
enclosed by the evolution trajectory. This leads to the first order
correction. Furthermore, according to the quantum jump approach, the total
phase has a dynamical contribution even for $\theta =\pi /2$. These results
are obvious in contrast with the present study, in which the geometric phase
depends upon the ratio of the probabilities of the joint states with one and
zero excitations which remains unchanged during the system-environment
coupling. The geometric phase calculated through the kinematic method [26]
also contains a first order correction due to the loss of the
system-reservoir entanglement after tracing over the reservoir.

Now we turn to another model in which a two-level atom is driven by a
quantized field and coupled to a reservoir. In the frame rotating at the
atomic transition frequency, the system Hamiltonian is described by the JC
model
\begin{equation}
H_s=\delta a^{\dagger }a+g(a^{\dagger }\left| g\right\rangle \left\langle
e\right| +a\left| e\right\rangle \left\langle g\right| ),
\end{equation}
where $a^{\dagger }$ and $a$ are the creation and annihilation operators for
the quantized field, $g$ is the atom-field coupling constant, and $\delta $
is the detuning. Suppose that the quantum system is initially in the state $%
\left| \psi _0^s(0)\right\rangle =\left| e\right\rangle \left|
0\right\rangle $, where $\left| 0\right\rangle $ is the vacuum state of the
quantum field. After an interaction time $t$ the system+reservoir state can
be written in the form of Eq. (5), where
\begin{eqnarray}
\left| \psi _0^s(t)\right\rangle &=&\frac 1{2\lambda }e^{-(i\delta /2+\gamma
/4)t}\left\{ \left[ \left( \lambda +\frac \delta 2+i\frac \gamma 4\right)
e^{i\lambda t}\right. \right.  \nonumber \\
&&\ \ \ \left. +\left( \lambda -\frac \delta 2-i\frac \gamma 4\right)
e^{-i\lambda t}\right] \left| e\right\rangle \left| 0\right\rangle  \nonumber
\\
&&\ \ \ \left. -g\left( e^{i\lambda t}-e^{-i\lambda t}\right) \left|
g\right\rangle \left| 1\right\rangle \right\} ,
\end{eqnarray}
with $\lambda =\sqrt{g^2+\left( \delta /2+i\gamma /4\right) ^2}$. Then the
geometric phase of the total system is
\begin{equation}
\beta =-\delta t/2+\arg \left\{ \frac 1{2\lambda }\left[ \left( \lambda +%
\frac \delta 2+i\frac \gamma 4\right) e^{i\lambda t}+\left( \lambda -\frac
\delta 2-i\frac \gamma 4\right) e^{-i\lambda t}\right] \right\} .
\end{equation}
In the limit $\Omega \gg \gamma $, where $\Omega =\sqrt{g^2+\delta ^2/4}$,
we can expand $\lambda $ up to the second order in $\gamma /\Omega $: $%
\lambda \simeq \Omega +i\frac \delta {8\Omega }\gamma -g^2\gamma ^2/\left(
32\Omega ^4\right) $. After a full Rabi cycle with $T=\pi /\Omega $, the
geometric phase is $\beta \simeq \beta ^0+\pi \delta \left( 3g^2-\delta
^2/2\right) \gamma ^2/\left( 64\Omega ^5\right) $, where $\beta ^0=\pi
\left[ 1-\delta /(2\Omega )\right] $ is the geometric phase without
decoherence. This implies that the geometric phase is equal to $\beta ^0$ up
to the first order in $\gamma /\Omega $, reflecting its robustness against
dissipation. Based on previous approaches, the lowest order correction for
this open system is linear [37], which is in distinct contrast with the
present result. It should be noted that for this open system the geometric
phase calculated through the present approach is the only one that can be
directly measured in experiment [37].

This result can be understood in the following way. We can rewrite the total
system-reservoir Hamiltonian in the single-excitation subspace as

\begin{eqnarray}
H &=&\Omega \left( \left| +\right\rangle \left\langle +\right| -\left|
-\right\rangle \left\langle -\right| \right) +\sum_{k=1}^N\omega
_kb_k^{\dagger }b_k  \nonumber \\
&&+\sum_{k=1}^N\left\{ g_kb_k\left[ \cos (\theta /2)\left| +\right\rangle
+\sin (\theta /2)\left| -\right\rangle \right] \left\langle g,0\right|
+H.c.\right\} ,
\end{eqnarray}
where $\left| +\right\rangle =\cos (\theta /2)\left| e,0\right\rangle +\sin
(\theta /2)\left| g,1\right\rangle $ and $\left| -\right\rangle =\sin
(\theta /2)\left| e,0\right\rangle -\cos (\theta /2)\left| g,1\right\rangle $
are the dressed states of the Jaynes-Cummings system, where $\cos \theta
=\delta /(2\Omega )$. Under the condition that the Rabi frequency $\Omega $
is much larger than the atomic decay rate $\gamma $ (the linewidth of the
upper level), the probability for the system undergoing transition from one
dressed state to the other vanishes due to the high frequency evolution. In
other words, the dressed state $\left| +\right\rangle $ ($\left|
-\right\rangle $) can only emit a photon into the reservoir modes with the
frequency ranging from $\omega _a+\Omega -\gamma $ to $\omega _a+\Omega
+\gamma $ ($\omega _a-\Omega -\gamma $ to $\omega _a-\Omega +\gamma $). When
$\gamma \ll \Omega $, the spectra of these two dressed states do not
overlap, and they evolve independently in the reservoir, with the
corresponding decaying rates $\gamma _{+}=\gamma \cos ^2(\theta /2)$ and $%
\gamma _{-}=\gamma \sin ^2(\theta /2)$. Then the state evolution of the
whole system can be approximately written as
\begin{equation}
\left| \psi ^{sr}(t)\right\rangle \simeq e^{-i(\delta t+\varphi )/2}\left[
\cos (\theta /2)\left| e_{+}\right\rangle +e^{i\varphi }\sin (\theta
/2)\left| e_{-}\right\rangle \right] ,
\end{equation}
where $\left| e_{\pm }\right\rangle =e^{-\gamma _{\pm }t/2}\left| \pm
\right\rangle \left| \stackrel{-}{0}\right\rangle +\sqrt{1-e^{-\gamma _{\pm
}t}}\left| g,0\right\rangle \left| \stackrel{-}{1}_{\pm }\right\rangle $, $%
\varphi =2\Omega t$, and $\left| \stackrel{-}{1}_{\pm }\right\rangle
=\sum_{k=1}^N\lambda _{k,\pm }\left|
0_1...0_{k-1}1_k0_{k+1}...0_N\right\rangle $ are the normalized collective
reservoir states with $\left\langle \stackrel{-}{1}_{+}\right| \left.
\stackrel{-}{1}_{-}\right\rangle =0$. Similar to the previous case, the time
evolution of either of the states $\left| e_{+}\right\rangle $ and $\left|
e_{-}\right\rangle $ has no effect on the acquired geometric phase $\beta $.
Therefore, after a time $T=\pi /\Omega $ the acquired geometric phase $\beta
$ is also directly related to the solid angle swept by the vector pointing ($%
\theta $, $\varphi $) on the Bloch sphere.

The approach can be generalized to the case when the system is composed of
different subsystems, each coupled to an independent reservoir. As an
example, we consider the dissipative JC model in which the atomic
spontaneous emission and photon decay rates are $\gamma $ and $\kappa $,
respectively. Suppose that the system is initially in the state $\left|
e\right\rangle \left| n\right\rangle $. Then the evolution of the state
component with no atomic spontaneous decay and photon loss is
\begin{eqnarray}
\left| \psi _0^s(t)\right\rangle &=&\frac 1{2\lambda _n}e^{-[(i\delta
+\kappa /2)(2n+1)/2+\gamma /4]t}\left\{ \left[ \left( \lambda _n+\frac \delta
2+i\frac{\gamma -\kappa }4\right) e^{i\lambda _nt}\right. \right.  \nonumber
\\
&&\ \left. +\left( \lambda _n-\frac \delta 2-i\frac{\gamma -\kappa }4\right)
e^{-i\lambda _nt}\right] \left| e\right\rangle \left| n\right\rangle
\nonumber \\
&&\ \left. -g\sqrt{n+1}\left( e^{i\lambda _nt}-e^{-i\lambda _nt}\right)
\left| g\right\rangle \left| n+1\right\rangle \right\} ,
\end{eqnarray}
where $\lambda _n=\sqrt{g^2(n+1)+\left[ \delta /2+i(\gamma -\kappa
)/4\right] ^2}$. When $(\gamma -\kappa )/\Omega _n\ll 1$ with $\Omega _n=%
\sqrt{g^2(n+1)+\delta ^2/4}$, up to the second order in $(\gamma -\kappa
)/\Omega _n$ we have $\lambda _n\simeq \Omega _n+i\frac \delta {8\Omega _n}%
(\gamma -\kappa )-g^2(\gamma -\kappa )^2/\left( 32\Omega ^4\right) $. After
a time $T=\pi /\Omega _n$, the acquired geometric phase is $\beta _n\simeq
\beta _n^0+\pi \delta \left( 3g^2-\delta ^2/2\right) (\gamma -\kappa
)^2/\left( 64\Omega ^5\right) $, where $\beta _n^0=\pi \left[ 1-\delta
/(2\Omega _n)\right] $ is the result without decoherence. This implies that
to the first order correction $\beta _n$ is affected neither by the atomic
spontaneous decay nor by the photon loss. This phase can also be directly
measured in the Ramsey interference experiment by employing an auxiliary
state $\left| f\right\rangle $ that is not coupled to the quantized field
[37].

In conclusion, we have described a new method for calculating the geometric
phases of a quantum system subjected to decoherence based on joint
system-reservoir evolution. This method requires neither to measure the
quantum jumps nor to trace over the reservoir modes, offering a way to get
around the nonunitary evolution obstacle. The geometric phase is associated
with the curvature of the total system-reservoir Hilbert space and always
can be well defined. In particular, we have shown that the geometric phase
can be calculated without completely knowing the state evolution of the
overall system. Using this method, we have calculated the geometric phase of
a qubit evolving under different Hamiltonians and subject to decoherence,
and show that for the dispersive system Hamiltonian it is completely
insensitive to decoherence. In case of low docoherence, the correction of
this phase associated with the JC evolution vanishes up to the first order
correction. These results are completely different from previous works,
which showed that for nonadiabatic evolution the decoherence effect appears
in the first order correction. We have presented a geometric explanation to
this difference. Furthermore, we have shown that for the dissipative JC
model the geometric phase calculated through the present method is the only
one that can be directly detected in experiment. Our method is generic and
can be applied to various physical systems subject to spontaneous decay.

This work was supported by the National Natural Science Foundation of China
under Grant No. 11374054 and the Major State Basic Research Development
Program of China under Grant No. 2012CB921601.

\section{Supplementary information: Open system geometric phase
based on system-reservoir joint state evolution}

In the supplementary material we show how to measure the open system
geometric phase calculated through our approach in experiment.

We here show how the geometric phase associated with the joint
system-reservoir state evolution can be measured in interference
experiments. When the system is initially in the state $\left| \psi
_0^s(0)\right\rangle $, after a cyclic evolution the
system-reservoir joint state can be written as
\begin{equation}
\eta e^{i\phi }\left| \psi _0^s(0)\right\rangle \left| \stackrel{-}{0}%
\right\rangle +\sqrt{1-\eta ^2}\left| \psi ^{^{\prime
}sr}\right\rangle ,
\end{equation}
where $\left| \psi ^{^{\prime }sr}\right\rangle $ is the joint state
component that is orthogonal to $\left| \psi _0^s(0)\right\rangle
\left|
\stackrel{-}{0}\right\rangle $, $\eta $ is a positive and real number, and $%
\phi $ is the total phase obtained by the system+reservoir state. To
measure this phase, we should use a system reference state $\left|
a\right\rangle $ that does not evolve. Suppose the system is
initially prepared in the superposition state $\left[ \left| \psi
_0^s(0)\right\rangle +\left| a\right\rangle \right] /\sqrt{2}$.
After the cyclic evolution, the system+reservoir state is
\begin{equation}
\frac 1{\sqrt{2}}\left\{ \left[ \eta e^{i\phi }\left| \psi
_0^s(0)\right\rangle +\left| a\right\rangle \right] \left| \stackrel{-}{0}%
\right\rangle +\sqrt{1-\eta ^2}\left| \psi ^{^{\prime
}sr}\right\rangle \right\} .
\end{equation}
By this way the total phase $\phi $ is encoded in the relative
probability amplitude associated with the states $\left| \psi
_0^s(0)\right\rangle $ and $\left| a\right\rangle $. When the
dynamical phase is zero, the geometric phase $\beta $ can be
directly measured by the relevant interference experiment.

Let us illustrate the procedure for measuring this geometric phase
with a typical example. Consider the dissipative Jaynes-Cummings
model in which a two-level atom is driven by a quantized field and
coupled to a reservoir. The system Hamiltonian $H_s$ for this model
is given by Eq. (10) of the main text. We aim to measure the
geometric phase associated with the evolution of
the initial state $\left| \psi _0^s(0)\right\rangle \left| \stackrel{-}{0}%
\right\rangle =\left| e,0\right\rangle \left| \stackrel{-}{0}\right\rangle $%
. Since $\left\langle \psi _0^s(0)\right| H_s\left| \psi
_0^s(0)\right\rangle =0$, no dynamical phase is accumulated during
the evolution. The ground state $\left| g,0\right\rangle $ does not
evolve and hence can be used as the reference state. The system is
initially prepared in the state $(\left| e\right\rangle +\left|
g\right\rangle )\left| 0\right\rangle /\sqrt{2}$. In the limit
$\Omega =\sqrt{g^2+\delta ^2/4}\gg \gamma $, after a Rabi cycle
($T=\pi /\Omega $) the system-reservoir state approximately evolves
to
\begin{eqnarray}
\left| \psi ^{sr}(T)\right\rangle  &\simeq &\frac
1{\sqrt{2}}e^{i\beta }\left( u\left| e,0\right\rangle +v\left|
g,1\right\rangle \right) \left|
\stackrel{-}{0}\right\rangle +  \nonumber \\
&&+\xi \left| g,0\right\rangle \left| \stackrel{-}{1}\right\rangle +\frac 1{%
\sqrt{2}}\left| g,0\right\rangle \left| \stackrel{-}{0}\right\rangle
,
\end{eqnarray}
where $u=1-\frac 14\gamma T\left( 1+\cos ^2\theta \right) $, $v=\frac 18%
\gamma T\sin (2\theta )$, $\xi =\frac 1{\sqrt{2}}\sqrt{1-u^2-v^2}$,
$\cos \theta =\delta /(2\Omega )$. Then we perform the
transformation on the atom: $\left| g\right\rangle \rightarrow
(\left| g\right\rangle -\left| e\right\rangle )/\sqrt{2}$ and
$\left| e\right\rangle \rightarrow (\left| e\right\rangle +\left|
g\right\rangle )/\sqrt{2}$, leading to
\begin{eqnarray}
\left| \psi ^{^{\prime }sr}(T)\right\rangle  &\simeq &\frac 12\left[
\left( 1+ue^{i\beta }\right) \left| g\right\rangle +\left(
ue^{i\beta }-1\right)
\left| e\right\rangle \right] \left| 0\right\rangle \left| \stackrel{-}{0}%
\right\rangle   \nonumber \\
&&+\frac 1{\sqrt{2}}\left( \left| g\right\rangle -\left|
e\right\rangle \right) \left( \frac 1{\sqrt{2}}ve^{i\beta }\left|
1\right\rangle \left|
\stackrel{-}{0}\right\rangle +\xi \left| 0\right\rangle \left| \stackrel{-}{1%
}\right\rangle \right) .
\end{eqnarray}
The final probability for detecting the atom in the state $\left|
g\right\rangle $ is $P_g=\frac 12(1+u\cos \beta )$. Therefore, we
have $\cos \beta =\left( 2P_g-1\right) /u$, which implies that the
geometric phase based on our approach can be directly measured
through the Ramsey interference experiment [1]. Here we have assumed
that there is only one decaying channel for the excited state
$\left| e\right\rangle $. In the presence of more than one lower
states and hence decaying channels, the
geometric phase remains unchanged, while the probability $P_g$ becomes $P_g=%
\frac 14[(1+\gamma _g/\gamma )+\left( u^2+v^2\right) (1-\gamma _g/\gamma )]+%
\frac 12u\cos \beta $, where $\gamma _g$ is the spontaneous emission
rate for decaying to $\left| g\right\rangle $. It should be noted
that according to the previous approaches [2,3] the total phase for
this open system contains a dynamical contribution $\phi _d\simeq
-\pi ^2g^2\delta \gamma /(8\Omega ^4)$, and hence the corresponding
geometric phase cannot be directly measured in experiment. In other
words, with the previous definitions the phase measured in the
Ramsey interference contains a dynamical component. Based on these
method, the correction of the geometric phase due to decoherence is
$\pi ^2g^2\delta \gamma /(8\Omega ^4)$ up to the first order in
$\gamma /\Omega $.

For the case that the quantized field initially contains $n$ photons
and the
cavity dissipation is also considered, we should use an auxiliary state $%
\left| f\right\rangle $ that is not coupled to the quantized field
to measure the acquired geometric phase. The states $\left|
g\right\rangle $ and $\left| f\right\rangle $ can be chosen to be
two hyperfine levels of the atomic ground state. The system is
initially prepared in the state $(\left| e\right\rangle +\left|
f\right\rangle )\left| n\right\rangle /\sqrt{2}$.
Under the condition $\Omega _n=\sqrt{g^2(n+1)+\delta ^2/4}\gg \gamma $, $%
(n+1)\kappa $, after a Rabi cycle ($T=\pi /\Omega _n$) the
system-reservoir state approximates
\begin{eqnarray}
\left| \psi ^{sr}(T)\right\rangle &\simeq &\frac
1{\sqrt{2}}e^{i\beta _n}\left( u_n\left| e,n\right\rangle +v_n\left|
g,n+1\right\rangle \right) \left| \stackrel{-}{0}\right\rangle
_a\left| \stackrel{-}{0}\right\rangle _p+
\nonumber \\
&&\ +\xi _n\left[ \frac{p_n}{\sqrt{2}}\left( \left| g,n\right\rangle
\left|
\stackrel{-}{1}_g\right\rangle _a+\left| f,n\right\rangle \left| \stackrel{-%
}{1}_f\right\rangle _a\right) \left| \stackrel{-}{0}\right\rangle
_p\right.
\nonumber \\
&&\left. +\left( q_n\left| e,n-1\right\rangle +s_n\left|
g,n\right\rangle
\right) \left| \stackrel{-}{0}\right\rangle _a\left| \stackrel{-}{1}%
\right\rangle _p\right] +\frac 1{\sqrt{2}}e^{-n\kappa T/2}\left|
f,n\right\rangle \left| \stackrel{-}{0}\right\rangle _a\left| \stackrel{-}{0}%
\right\rangle _p  \nonumber \\
&&+\frac 1{\sqrt{2}}\sqrt{1-e^{-n\kappa T}}\left| f,n-1\right\rangle
\left| \stackrel{-}{0}\right\rangle _a\left|
\stackrel{-}{1}\right\rangle _p,
\end{eqnarray}
where $u_n=1-\frac 14\left[ (2n+1)\kappa +\gamma +(\gamma -\kappa
)\cos
^2\theta _n\right] T$, $v_n=\frac 18(\gamma -\kappa )T\sin (2\theta _n)$, $%
\xi _n=\frac 1{\sqrt{2}}\sqrt{1-u_n^2-v_n^2}$,
$p_n=\sqrt{\frac{\gamma \left( 1+\cos ^2\theta _n\right) }{\gamma
\left( 1+\cos ^2\theta _n\right) +\kappa (2n+\sin ^2\theta _n)}}$,
$q_n=\sqrt{\frac{n\kappa \left( 1+\cos ^2\theta _n\right) }{\gamma
\left( 1+\cos ^2\theta _n\right) +\kappa
(2n+\sin ^2\theta _n)}}$, $s_n=\sqrt{\frac{(n+1)\kappa \sin ^2\theta _n}{%
\gamma \left( 1+\cos ^2\theta _n\right) +\kappa (2n+\sin ^2\theta _n)}}$, $%
\cos \theta _n=\delta /\left( 2\Omega _n\right) $, the subscripts
$a$ and $p$ label the reservoirs to which the atom and quantized
photonic field are
coupled, respectively, and $\left| \stackrel{-}{1}_j\right\rangle _a$ ($%
j=g,f $) denotes the state of reservoir of the atom associated with
decaying to $\left| j\right\rangle $. We here have discarded the
trivial common phase factor $e^{-in\delta T}$ and assumed that the
spontaneous emission rates for
decaying to $\left| g\right\rangle $ and $\left| f\right\rangle $ are both $%
\gamma /2$. The subsequent transformations $\left| f\right\rangle
\rightarrow (\left| f\right\rangle -\left| e\right\rangle )/\sqrt{2}$ and $%
\left| e\right\rangle \rightarrow (\left| e\right\rangle +\left|
f\right\rangle )/\sqrt{2}$ leading to
\begin{eqnarray}
\left| \psi ^{^{\prime }sr}(T)\right\rangle &\simeq &\frac 12\left[
\left( e^{-n\kappa T/2}+e^{i\beta _n}u_n\right) \left|
f,n\right\rangle +\left( e^{i\beta _n}u_n-e^{-n\kappa T/2}\right)
\left| e,n\right\rangle \right.
\nonumber \\
&&\left. +\sqrt{2}v_n\left| g,n+1\right\rangle \right] \left| \stackrel{-}{0}%
\right\rangle _a\left| \stackrel{-}{0}\right\rangle _p  \nonumber \\
&&\ \ +\xi _n\left\{ \frac{p_n}{\sqrt{2}}\left[ \left|
g,n\right\rangle \left| \stackrel{-}{1}_g\right\rangle _a+\frac
1{\sqrt{2}}(\left| f\right\rangle -\left| e\right\rangle )\left|
n\right\rangle \left| \stackrel{-}{1}_f\right\rangle _a\right]
\left| \stackrel{-}{0}\right\rangle
_p\right.  \nonumber \\
&&\left. +\left[ \frac 1{\sqrt{2}}q_n(\left| e\right\rangle +\left|
f\right\rangle )\left| n-1\right\rangle +s_n\left| g,n\right\rangle
\right] \left| \stackrel{-}{0}\right\rangle _a\left|
\stackrel{-}{1}\right\rangle
_p\right\}  \nonumber \\
&&\ +\frac 12\sqrt{1-e^{-n\kappa T}}(\left| f\right\rangle -\left|
e\right\rangle )\left| n-1\right\rangle \left|
\stackrel{-}{0}\right\rangle _a\left| \stackrel{-}{1}\right\rangle
_p.
\end{eqnarray}
In this case the final probability for detecting the atom in the state $%
\left| f\right\rangle $ is directly related to the geometric phase: $P_f=%
\frac 14\left[ 1+u_n^2+\xi _n^2\left( p_n^2+2q_n^2\right) \right] +\frac 12%
u_ne^{-n\kappa T/2}\cos \beta $. According to the previous
approaches [2,3], for this open system the measured phase contains a
dynamical component $\phi _d\simeq -\pi ^2g^2\delta (n+1)(\gamma
-\kappa )/(8\Omega ^4)$.


\begin{references}
\bibitem{1}  M. V. Berry, Proc. R. Soc. Lond. A 392, 45 (1984).

\bibitem{2}  J. Anandan, Nature 360, 307 (1992).

\bibitem{3}  A. Shapere and F. Wilczek, Geometric Phases in Physics (World
Scientific, Singapore, 1989).

\bibitem{4}  Y. Aharonov and J. Anandan, Phys. Rev. Lett. 58, 1593 (1987).

\bibitem{5}  S. Pancharatnam, Proc. Indian Acad. Sci. Sect. A 44, 247 (1956).

\bibitem{6}  J. Samuel and R. Bhandari, Phys. Rev. Lett. 60, 2339 (1988).

\bibitem{7}  A. Uhlmann, Rep. Math. Phys. 24, 229 (1986).

\bibitem{8}  E. Sj\"oqvist et al., Phys. Rev. Lett. 85, 2845 (2000).

\bibitem{9}  P. Zanardi and M. Rasetti, Phys. Lett. A 264, 94 (1999).

\bibitem{10}  J.A. Jones, V. Vedral, A. Ekert, and G. Castagnoli, Nature 403,
869 (2000).

\bibitem{11}  G. Falci, R. Fazio, G.M. Palma, J. Siewert, and V. Vedral, Naute
407, 355 (2000).

\bibitem{12}  L.M. Duan, J.I. Cirac, and P. Zoller, Science 292, 1695 (2001).

\bibitem{13}  X.B. Wang and M. Keiji, Phys. Rev. Lett. 87, 097901 (2001).

\bibitem{14}  S.L. Zhu and Z. D. Wang, Phys. Rev. Lett. 89, 097902 (2002).

\bibitem{15}  D. Leibfried et al., Nature 422, 412 (2003).

\bibitem{16}  S.B.\ Zheng, Phys. Rev. A 70, 052320 (2004).

\bibitem{17}  G. De Chiara and G. M. Palma, Phys. Rev. Lett. 91, 090404 (2003).

\bibitem{18}  A. Nazir, T. Spiller, and W. J. Munro, Phys. Rev. A 65, 042303
(2002).

\bibitem{19}  A. Blais and A.-M. S. Tremblay, Phys. Rev. A 67, 012308 (2003).

\bibitem{20}  P. Solinas, M. Sassetti, P. Truini, and N. Zangh , New J. Phys.
14, 093006 (2012).

\bibitem{21}  P. J. Leek et al., Science 318, 1889 (2007).

\bibitem{22}  S. Filipp, J. Klepp, Y. Hasegawa, C. Plonka-Spehr, U. Schmidt,
P. Geltenbort, and H. Rauch, Phys. Rev. Lett. 102, 030404 (2009).

\bibitem{23}  S. Berger, M. Pechal, A. A. Abdumalikov Jr., C. Eichler, L.
Steffen, A. Fedorov, A. Wallraff, S. Filipp, Phys. Rev. A 87, 060303(R)
(2013).

\bibitem{24}  S.B. Zheng, Phys. Rev. A 85, 022128 (2012).

\bibitem{25}  M. Ericsson, E. Sj\"oqvist, J. Br\"annlund, D. K. L. Oi, and A.
K. Pati, Phys. Rev. A 67, 020101(R) (2003).

\bibitem{26}  D. M. Tong, E. Sjoqvist, L. C. Kwek, and C. H. Oh, Phys. Rev.
Lett. 93, 080405 (2004).

\bibitem{27}  A. Carollo, I. Fuentes-Guridi, M. F. Santos, and V. Vedral,
Phys. Rev. Lett. 90, 160402 (2003).

\bibitem{28}  A. Carollo, I. Fuentes-Guridi, M. F. Santos, and V. Vedral,
Phys. Rev. Lett. 92, 020402 (2004).

\bibitem{29}  R. S. Whitney and Y. Gefen, Phys. Rev. Lett. 90, 190402 (2003).

\bibitem{30}  R. S. Whitney, Y. Makhlin, A. Shnirman, and Y. Gefen, Phys. Rev.
Lett. 94, 070407 (2005).

\bibitem{31}  X.X. Yi, D. M. Tong, L.C. Wang, L.C. Kwek, and C.H. Oh, Phys.
Rev. A 73, 052103 (2006)

\bibitem{32}  P. Solinas, M. Mottonen, J. Salmilehto, and J. P. Pekola, Phys.
Rev. B 82, 134517 (2010).

\bibitem{33}  F.C. Lombardo, P.I. Villar, Phys. Rev. A 81, 022115 (2010).

\bibitem{34}  P.I. Villar, and F.C. Lombardo, Phys. Rev. A 83, 052121 (2011).

\bibitem{35}  F.M. Cucchietti, J.-F. Zhang, F.C. Lombardo, P.I. Villar, R.
Laflamme, Phys. Rev. Lett. 105, 240406 (2010).

\bibitem{36}  M. Ericsson, A.K. Pati, E. Sj\"oqvist1, J. Br\"annlund, and
D.K.L. Oi, Phys. Rev. Lett. 91, 090405 (2003).

\bibitem{37}  see supplementary material.

\bibitem{38}  M.O. Scully and M.S. Zubairy, {\em Quantum Optics} (Cambridge
University Press, Cambridge, 1997).
\end{references}

\begin{references}
\bibitem{1}  P. Bertet, S. Osnaghi, A. Rauschenbeutel, G. Nogues, A.
Auffeves, M. Brune, J.M. Raimond, and S. Haroche, Nature 411, 166
(2001).

\bibitem{2}  D. M. Tong, E. Sjoqvist, L. C. Kwek, and C. H. Oh, Phys. Rev.
Lett. 93, 080405 (2004).

\bibitem{3}  A. Carollo, I. Fuentes-Guridi, M. F. Santos, and V. Vedral,
Phys. Rev. Lett. 90, 160402 (2003).
\end{references}
\end{document}